\documentclass[reprint,groupedaddress,showpacs,amsmath,amssymb,aps,pra,floatfix]{revtex4-1}
\usepackage{graphicx}
\usepackage{xcolor,colortbl}
\usepackage{array}
\usepackage{bm}

\newcommand{\rrr}[1]{\vskip 0.2cm \noindent{\it #1} ---}

\begin{document}

\title{Schrödinger's cat state of optical parallel universes}

\author{Yu-Qing Cui${}^{1}$}

\author{Tian-Ming Zhao${}^{1}$}
\email{zhaotm@scnu.edu.cn}  

\author{Rong-Xin Miao${}^{2}$}
\email{miaorx@mail.sysu.edu.cn}

\author{Jin-Dong Wang${}^{3}$}

\author{Huanyang Chen${}^{4}$}

\affiliation{${}^{1}$Guangdong Provincial Key Laboratory of Nanophotonic Functional Materials and Devices, South China Normal University, Guangzhou, 510631, China\\
${}^{2}$School of Physics and Astronomy, Sun Yat-Sen University, Zhuhai, 519082, China\\
${}^{3}$Guangdong Provincial Key Laboratory of Quantum Engineering and Quantum Materials, South China Normal University, Guangzhou, 510631, China\\
${}^{4}$Department of Physics, Xiamen University, Xiamen 361005, China}


\begin{abstract}
Parallel worlds are imaginative ideas in quantum mechanics and cosmology. The superpositions of parallel worlds are novel states of quantum gravity and have no classical correspondences generally. In this letter, we investigate the superposition or the Schrödinger's cat state of optical parallel worlds, which could be realized in the laboratory and may shed some light on the detection of parallel universes in the real world.  We propose two realizable experimental schemes that enable exploring the mysterious `parallel universes'  by a Mach-Zehnder interferometer. The first one is based on an atomic ensemble in a superposition state, which is a fat Schrödinger's cat state. The second one is to prepare a photon in a superposition of different paths, where each path lies in an optical parallel universe.  
\end{abstract}
  

\maketitle

The parallel worlds are intriguing and imaginative ideas in quantum physics, which are also popular science fiction topics. There are several different kinds of parallel worlds. The famous one is the many world interpretation of quantum mechanics \cite{Everett}, which conjectures all the outcomes that can occur actually do happen, but only one outcome can happen in each universe. Roughly speaking, during the evolution of the Schr\"odinger equation, the world splits into many different branches, and our universe is just one of them \cite{Tegmark:2007wh}. The second kind of parallel world is the multiverse of eternal inflation \cite{Eternalinflation}, where the space is broken up into infinite causally disconnected bubble universes. In an eternally inflating universe, anything that can happen will happen \cite{Guth}. Interestingly, the many worlds of quantum mechanics and the many worlds of the multiverse are conjectured to be equivalent \cite{Nomura:2011dt,Bousso:2011up}. For other types of parallel universes, please refer to \cite{Tegmark:2003sr}. 

Although novel, parallel worlds are difficult to be observed in experiments. See \cite{Feeney:2010dd,Feeney:2010jj} for some observational tests of multiverses. Thus, it is interesting to study the optical analogy of parallel worlds, which is much easier to be realized in experiments and may shed some light on the detection of parallel universes in the real world.  Thanks to the rapid developments of metamaterials and transformation optics \cite{Leonhardt,Pendry}, the optical mimicking of novel spacetime such as black holes \cite{Chen:2009mk}, worm holes \cite{Greenleaf:2007bd}, de Sitter universes \cite{Li:2009pm,Li:2009zy}, multiverses \cite{Smolyaninov:2010mc} and other geometries \cite{Leonhardt:2006ai,Zhang2009,Chen:2020dxy,Zhao:2011nq} are possible in laboratory. However, since metamaterials are classical objects, these works can only mimic the classical spacetime. In this paper, we explore the quantum effects: the superposition or the Schrödinger's cat state of optical parallel worlds. 

 \begin{figure}[t]
\centering
\includegraphics[width=7.9cm]{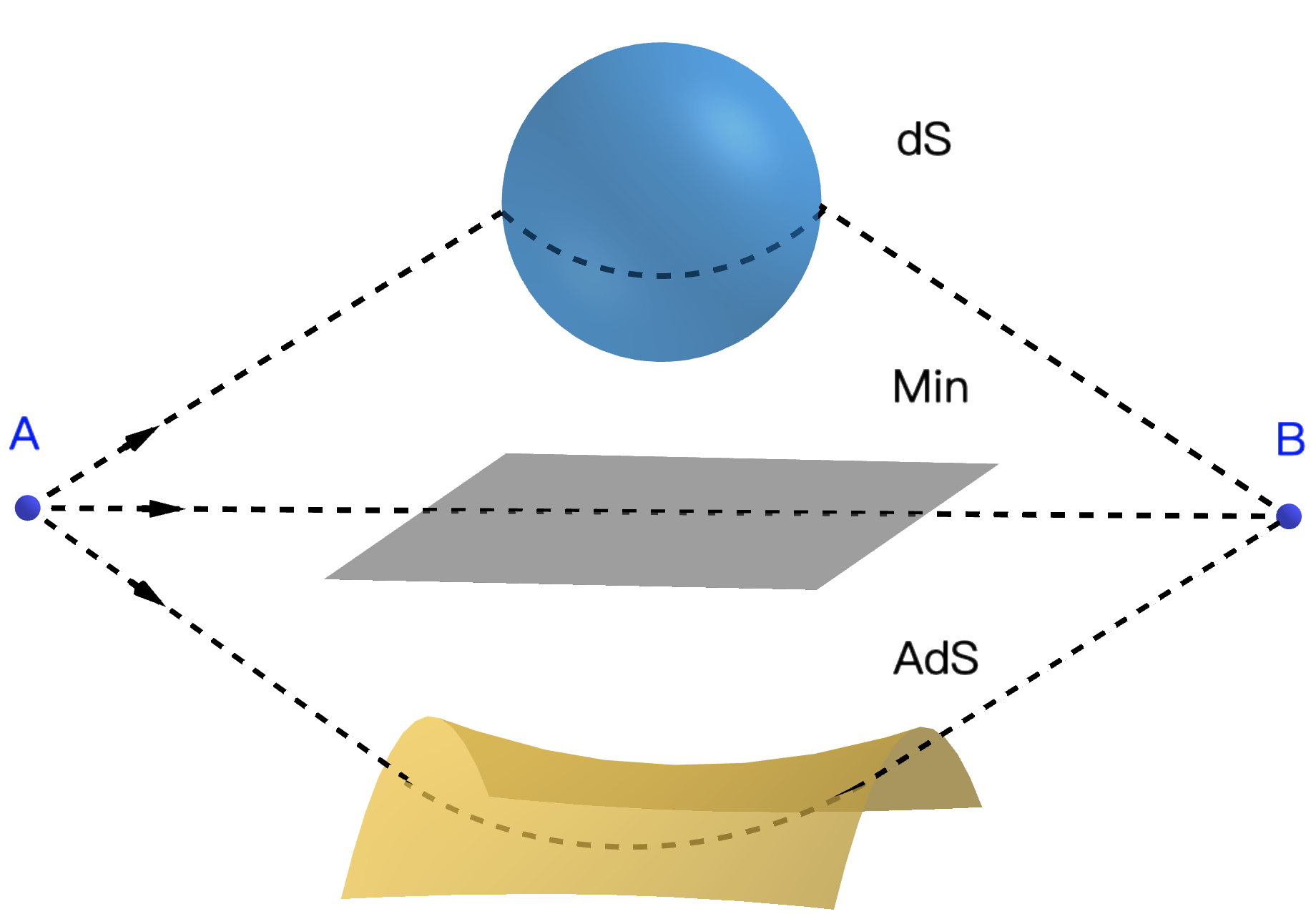}
\caption{Parallel worlds of ants. Due to the quantum fluctuations of spacetime, the ant travels in infinite parallel worlds before it arrives at B. Here dS (sphere), Min (plane) and AdS (saddle) denote the maximum symmetrical spaces with positive, zero and negative curvatures.}
\label{ant}
\end{figure}

Consider two quantum states of gravity which correspond to two kinds of universes. Without loss of generality, we take  Minkowski spacetime (Min) $|\Psi_{\text{Min}}\rangle$ and de Sitter universe (dS) $|\Psi_{\text{dS}}\rangle$ as examples. Since the expansion of our universe is accelerating currently \cite{SupernovaSearchTeam:1998fmf,SupernovaCosmologyProject:1998vns}, dS is a good approximation of our universe at cosmic scale. On the other hand,  Min is a good approximation of our universe at small scale.  According to the standard quantum mechanics, the superposition of the two states
\begin{eqnarray}\label{superposition}
|\Psi\rangle=\frac{1}{\sqrt{2}} \Big(|\Psi_{\text{Min}}\rangle+ |\Psi_{\text{dS}}\rangle\Big),
\end{eqnarray}
is also a quantum state of gravity. However, unlike  $|\Psi_{\text{Min}}\rangle$ and $|\Psi_{\text{dS}}\rangle$, the state (\ref{superposition}) has no classical correspondence. Obviously, one cannot live in two different classical universes at the same time. Actually (\ref{superposition}) is an analogy of Schrödinger's cat, where the poor cat is replaced by the spacetime.  Another way to understand this is as follows. The classical spacetime is a solution to Einstein equations, which are highly non-linear equations. The linear combinations of two solutions cannot be a solution to non-linear equations. As a result, (\ref{superposition}) cannot be classical spacetime. Instead, it is a non-trivial state of quantum gravity. The main purpose of this letter is to study this novel quantum state of parallel worlds in optical systems.

Note that (\ref{superposition}) is the superposition of parallel worlds before a measurement is performed. According to quantum mechanics, although it is difficult to observe, such a superposition state does exist in a real world if gravity is quantum \cite{Marletto,Foo:2020jmi}. Suppose that an ant crawls from point A to point B and no observation is made in this process. Due to the quantum fluctuations of spacetime, the ant travels in infinite parallel worlds before it arrives at B. See Fig.\ref{ant} for an example, where parallel worlds are labelled by dS, Min and anti de Sitter space (AdS). At the end point B, the ant retains the `memory' of parallel worlds in the phases, which can be extracted by interference experiments in principle. In this letter, the ant is replaced by a photon and the many worlds of spacetime are mimicked by optical parallel universes.

\rrr{Mimicking spacetime}  To start, let us give a quick review on how to mimic curved spacetime in an optical system. Without loss of generality, we take dS as an example. dS describes a universe with a constant rate of accelerated expansion. In static coordinates, the metric is given by
\begin{eqnarray}\label{dSmetric0}
ds^2=- f(\bar{r}) c^2dt^2+\frac{d\bar{r}^2}{ f(\bar{r})}+\bar{r}^2d\Omega^2,
\end{eqnarray}
where $f(\bar{r})=(1-H^2\bar{r}^2)$, $H$ is the Hubble parameter, $c$ is the velocity of light, $d\Omega^2=d\theta^2+\sin^2\theta d\phi^2$ is the line element of unite sphere. 
According to \cite{Leonhardt:2006ai,Plebanski}, the curved spacetime with the metric $ds^2=-g_{00}c^2dt^2+g_{ij} dx^i dx^j$ can be mimicked by a dielectric with the refractive index
\begin{eqnarray}\label{permittivity}
n^{ij}=\frac{\sqrt{-g}}{\sqrt{\gamma}}\frac{g^{ij}}{g_{00}},
\end{eqnarray}
where $g$ is the determinant of metric, $g^{ij}$ is the inverse metric, $\gamma$ labels the determinant of the physical metric in laboratory. For examples, $\gamma=1$ for Cartesian coordinates and $\gamma=r^4 \sin^2\theta$ for spherical coordinates. A photon cannot distinguish if it lives in curved spacetime or a dielectric, since by design, Maxwell's equations are the same in them.  

From (\ref{permittivity}), we notice that the refractive index of (\ref{dSmetric0}) is anisotropic,  which makes the experiment difficult. To get a isotropic refractive index, we perform the coordinate transformation $\bar{r}=\frac{r}{1+ (H^2r^2/4)}$ so that $dr/r=d\bar{r}/(\bar{r}\sqrt{f(\bar{r})})$ and the metric (\ref{dSmetric0}) becomes
\begin{eqnarray}\label{dSmetric}
ds^2=-f(\bar{r}) c^2dt^2+\frac{\bar{r}^2}{r^2}\Big( dr^2+r^2d\Omega^2 \Big),
\end{eqnarray}
where the spatial metric in the braces is the standard isotropic metric in spherical coordinates. As a result, the refractive index (\ref{permittivity}) becomes isotropic
\begin{eqnarray}\label{permittivitydS}
n_{\text{dS}}(r)=\frac{1}{1-\frac{1}{4 } H^2r^2}\approx 1+\frac{1}{4 } H^2r^2+ O(H^4).
\end{eqnarray}
For the convenient of readers, in Table.\ref{table1} we list the refractive indexes for various static spacetime \cite{Chen:2020dxy,note1} and for an expanding universe with Robertson-Walker metric
\begin{eqnarray}\label{RWmetric}
ds^{2}=-c^2dt^{2}+a^{2}(t)(dr^2+r^2d\Omega^2),
\end{eqnarray}
where $a(t)$ is the scale factor. When $a(t)=\exp(Ht)$, (\ref{RWmetric}) becomes the dS metric in time-dependent coordinates. 
\begin{table}[htbp]
	\centering 
	\caption{Refractive index of various spacetime}  
	\label{table1}  
	\begin{tabular}{|c|c|c|c|}  
		\hline  
		& & \\[-6pt]  		
		Spacetime& Metric & Refractive index\\ \hline
		& & \\[-6pt]  
		Min & $f=1$ & $n=1$ \\  \hline
		& & \\[-6pt]  
		 Black hole (BH) & $f=1-\frac{2M}{\bar{r}}$& $n=\frac{(M+2 r)^3}{4 r^2 (2 r-M)}$  \\  \hline
		& & \\[-6pt]  
		dS & $f=1-H^2\bar{r}^2$& $n=\frac{1}{1-\frac{1 }{4}H^2 r^2}$  \\\hline
		& & \\[-6pt]  
		dS-BH & $f=1-H^2\bar{r}^2-\frac{2M}{\bar{r}}$& $n\approx 1+\frac{H^2 r^2}{4 } +\frac{2M}{r}$  \\\hline
		& & \\[-6pt]  
		AdS & $f=1+H^2\bar{r}^2$& $n=\frac{1}{1+\frac{1 }{4} H^2r^2}$  \\\hline
		& & \\[-6pt]  
		AdS-BH & $f=1+H^2\bar{r}^2-\frac{2M}{\bar{r}}$& $n\approx 1-\frac{H^2 r^2}{4 } +\frac{2M}{r}$  \\\hline
		& & \\[-6pt]  
		Expanding universe & RW metric (\ref{RWmetric}) & $n=a(t)$  \\\hline
	\end{tabular}
\end{table}

\begin{figure}[t]
\centering
\includegraphics[width=6.8cm]{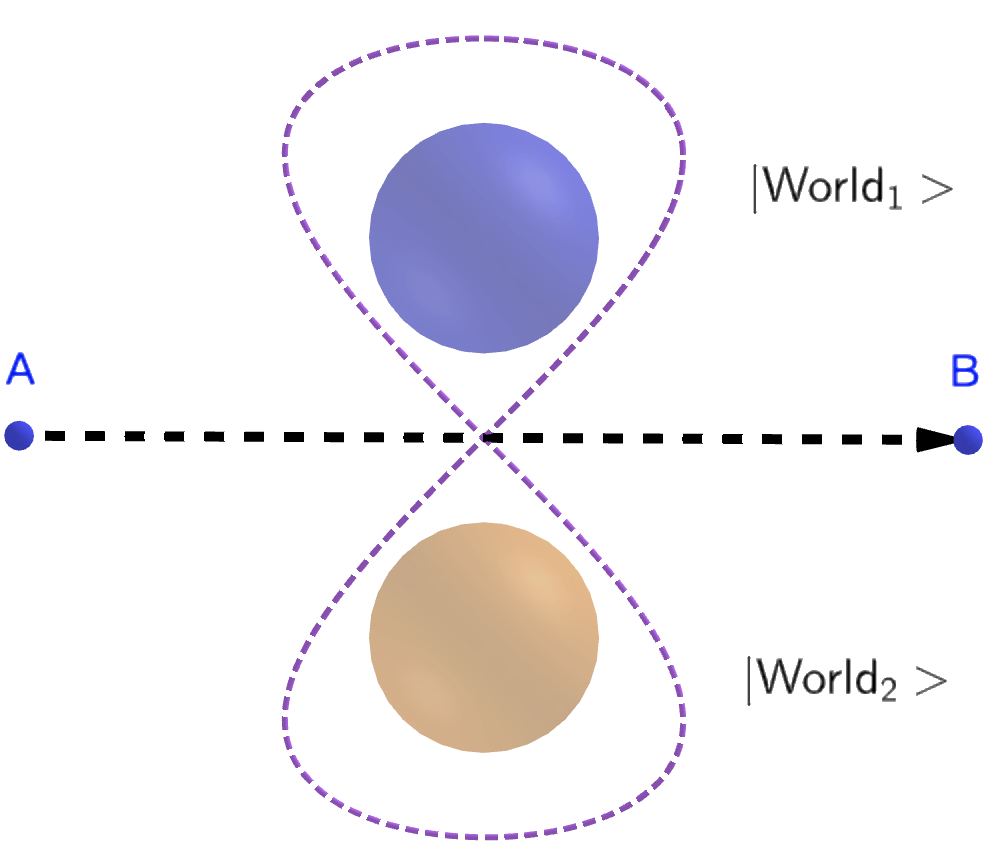}
\caption{A photon passes through a superposition state of (optical) parallel worlds $|\Psi\rangle=\Big(|\text{World}_1\rangle+ |\text{World}_2\rangle \Big)/\sqrt{2}$. 
}
\label{parallelworld1}
\end{figure}

\rrr{Scheme I}  Now we are ready to discuss the mimicking of parallel universes. We have two experimental  proposals. The first one is a perfect optical realization of the Schrödinger's cat state of parallel universes (\ref{superposition}), which is based on an atomic ensemble in a superposition state. See Fig.\ref{parallelworld1} for schematic. The external light can change the refractive index of an atomic ensemble. Suppose that the refractive index is sensitive to the polarization of external light, which can be realized by adding a polarizer in front of the atomic ensemble. Prepare the external light in a superposition of coherent
states $|\alpha_{x}\rangle$ and $|\alpha_{y}\rangle$, where $x,y$ denote the polarization directions. Let the atomic ensemble to be sensitive to only the y-polarization so that  $|\alpha_{y}\rangle$ changes the refractive index to be $n_2$, while $|\alpha_{x}\rangle$ keeps invariant the refractive index $n_1\approx 1$. Then the external light $|\alpha\rangle=\frac{1}{\sqrt{2}}(|\alpha_{x}\rangle+ |\alpha_{y}\rangle)$ makes the atomic ensemble in a superposition state
\begin{eqnarray}\label{superpositionworlds}
|n\rangle=\frac{1}{\sqrt{2}} \Big(|n_{1}\rangle+ |n_{2}\rangle\Big),
\end{eqnarray}
which is an optical realization of the novel state of quantum gravity (\ref{superposition}). Here $n_1$ and $n_2$ denote the refractive indexes of the two parallel worlds. 

Some comments are in order. {\bf 1}. The key point is to prepare an atomic ensemble in the superposition state (\ref{superpositionworlds}). The above scheme is just a suggestion.  Experimentalists may have more feasible method to produce it. For example, design a light switch in a superposition of on and off $\frac{1}{\sqrt{2}}(|\text{on}\rangle+ |\text{off}\rangle)$. Then an external light passing though the light switch can induce the state (\ref{superpositionworlds}).  {\bf 2}.  $n_1$ and $n_2$ need not to be the refractive indexes of Min and dS. Note that Min and dS are just examples used in (\ref{superposition}) to illustrate the main ideas of parallel worlds. They can be replaced by other spacetime listed in Table.\ref{table1}. {\bf 3}.  (\ref{superpositionworlds}) is actually a fat Schrödinger's cat state of the atomic ensemble.  The fat Schrödinger's cat state has been created for 430 atoms in experiments \cite{factcatatom}. There are also interesting proposals for the enlargement of the optical Schrödinger's cat states \cite{factcatoptic}.  {\bf 4}.  We don't have to worry about the decoherence. Recall that the parallel worlds of this letter are only effective for photons. For the atomic ensemble of size $1 cm$, the characteristic time for a photon passing through it is about $t\approx 30 ps$, which is extremely short. As a result, one only needs to keep the coherence time of atomic ensemble to be longer than $30 ps$, which is possible in laboratory.  {\bf 5}.  As shown in Fig.\ref{parallelworld1}, consider a photon passing through an atomic ensemble in the superposition state (\ref{superpositionworlds}). The probability to observe the photon at the end point B is $P=(1+\cos\Delta \phi)/2$, where $\Delta \phi$ denotes the phase difference caused by the different refractive indexes of the optical parallel worlds.

\begin{figure}[t]
\centering
\includegraphics[width=7.6cm]{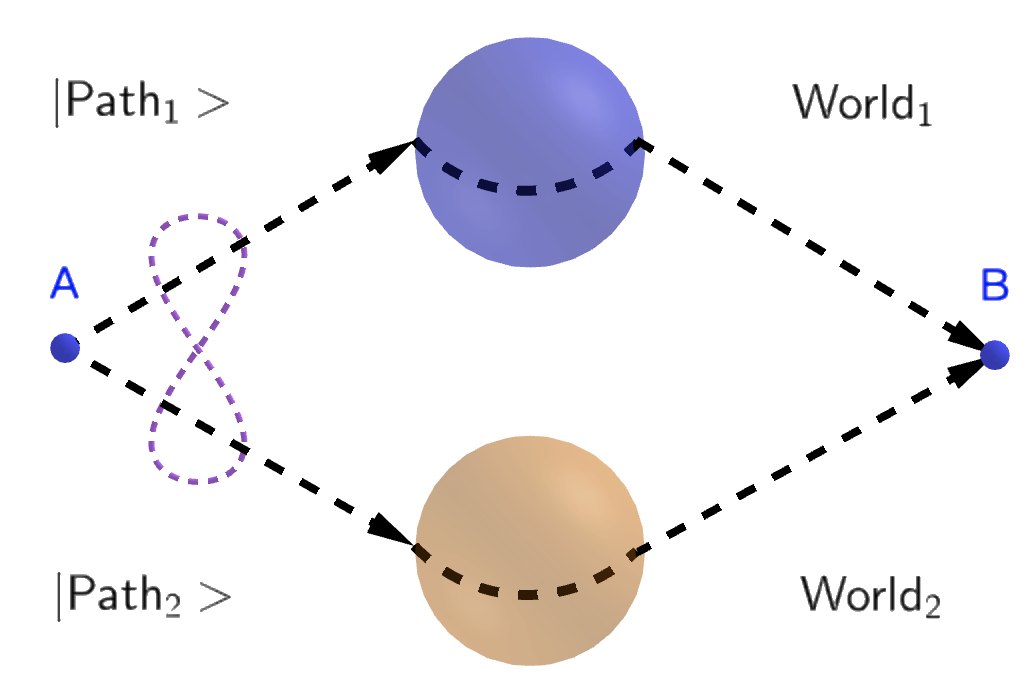}
\caption{A photon in a superposition state of paths $|\Psi'\rangle=\Big(|\text{Path}_1\rangle+ |\text{Path}_2\rangle\Big)/\sqrt{2}$ passes through two classical (optical) parallel worlds. 
}
\label{parallelworld2}
\end{figure}

\rrr{Scheme II} 
Let us go on to discuss the second scheme, which is a partial simulation of the novel state of parallel worlds (\ref{superposition}).  See Fig.\ref{parallelworld2}. Let a photon pass through two classical optical parallel worlds with equal probability amplitude. In other words, the photon is in a superposition state of the light path
\begin{eqnarray}\label{superpositionphoton}
|\Psi'\rangle=\frac{1}{\sqrt{2}} \Big(|\Psi'_{\text{pass Min}}\rangle+ |\Psi'_{\text{pass dS}}\rangle\Big).
\end{eqnarray}
From the viewpoint of photons, the optical parallel worlds are in the state (\ref{superposition}) effectively. In fact, a photon cannot distinguish scheme I and scheme II, that is because the probability to observe the photon at the end point B is exactly the same for scheme I and scheme II.  Note that, in scheme II, we only need to prepare a superposition state of a photon instead of an atomic ensemble, thus it can be achieved more easily in experiments.  

One can select arbitrary two kinds of spacetime of Table.\ref{table1} to constitute the parallel worlds.  The simplest choice is a Min together with a dS as in (\ref{superpositionphoton}), which is close to our real world: an expanding universe with small quantum fluctuations and small Hubble constant. To perform the experiments, one can make further simplifications.  Instead of making the difficult ball devices, we propose to make narrow cylinders and replace $n(r)$ of Table.\ref{table1} by $n(z)$, where z denotes the axial direction.  The photon propagating radially outward the isotropic universe can be mimicked well by the photon propagating along z-axial in the cylindrical optical device. That is because the s-wave is dominated as long as the optical length is not too large. 
If we focus on the s-wave, the narrow cylinder is a good approximation.


\rrr{Experimental design} 
Now let us present the exact experimental design. The experimental device is shown in Fig.\ref{P1}, where we lay a cylindrical vapor cell (parallel universe) along one of the light paths.  For scheme I, the atomic ensemble in the vapor cell is in a superposition state of refractive indexes $n_1$ and $n_2$ (\ref{superpositionworlds}), while for scheme II,  the refractive index of atomic ensemble is $n_2$. We set
\begin{eqnarray}\label{permittivityn1n2}
n_1= 1,\ \ n_2\approx 1+\frac{1 }{4}H^2 z^2,
\end{eqnarray}
which correspond to Min and dS respectively. Here $0\le z \le L$, $ HL<2 $, $L$ is the length of vapor cell and the Hubble parameter $H$ can be controlled by the intensity of external lights. The phase difference between the two  light paths of scheme II is given by
\begin{eqnarray}\label{phasedifference}
\Delta \phi=\frac{\pi }{6\lambda_0} H^2 L^3,
\end{eqnarray}
where $\lambda_0$ is the wavelength of photons in vacuum. Take $L=1cm$ and $\lambda_0=780nm$ as an example. By changing $HL$ from $0.01$ to $0.1$, the refractive index $n_2(L)$ varies from $1.000025$ to $1.0025$ and $\Delta \phi$ changes about $21 \pi$. The probabilities of obtaining a photon at the output ports $\pm$ are given by \cite{note2}
\begin{eqnarray}\label{probabilitiesI}
P_{I\pm}=\frac{1}{4} \left(2\pm\sqrt{2+2 \cos(\Delta \phi)}\right),
\end{eqnarray}
for scheme I, and
\begin{eqnarray}\label{probabilitiesII}
P_{II\pm}=\frac{1}{2}\Big(1\pm\cos(\Delta \phi)\Big),
\end{eqnarray}
for scheme II. 

\rrr{Atomic ensemble} 
Below we take the atomic ensemble as an example to illustrate the optical simulation of parallel worlds. Compared with solid metamaterials, it has the advantages of adjustable optical parameters and low absorption loss. In the past, the atomic ensemble is used to study the quantum coherence in three-level systems, such as electromagnetically induced transparency \cite{Fleischhauer:2005}, laser without inversion \cite{Harris:1989}, and refractive index enhancement without absorption \cite{Zibrov:1996}. These coherence studies have many applications, such as quantum memory \cite{Choi:2008}, precision magnetometry \cite{Scully:1992}, and high-frequency laser \cite{Scully:1994}. 
This research gives a full study on the elastic properties of atomic system parameters and realizes controllable refractive index of optical parallel worlds by the non-absorption index enhancement.

\begin{figure}[h]
	\includegraphics[width=0.8\columnwidth]{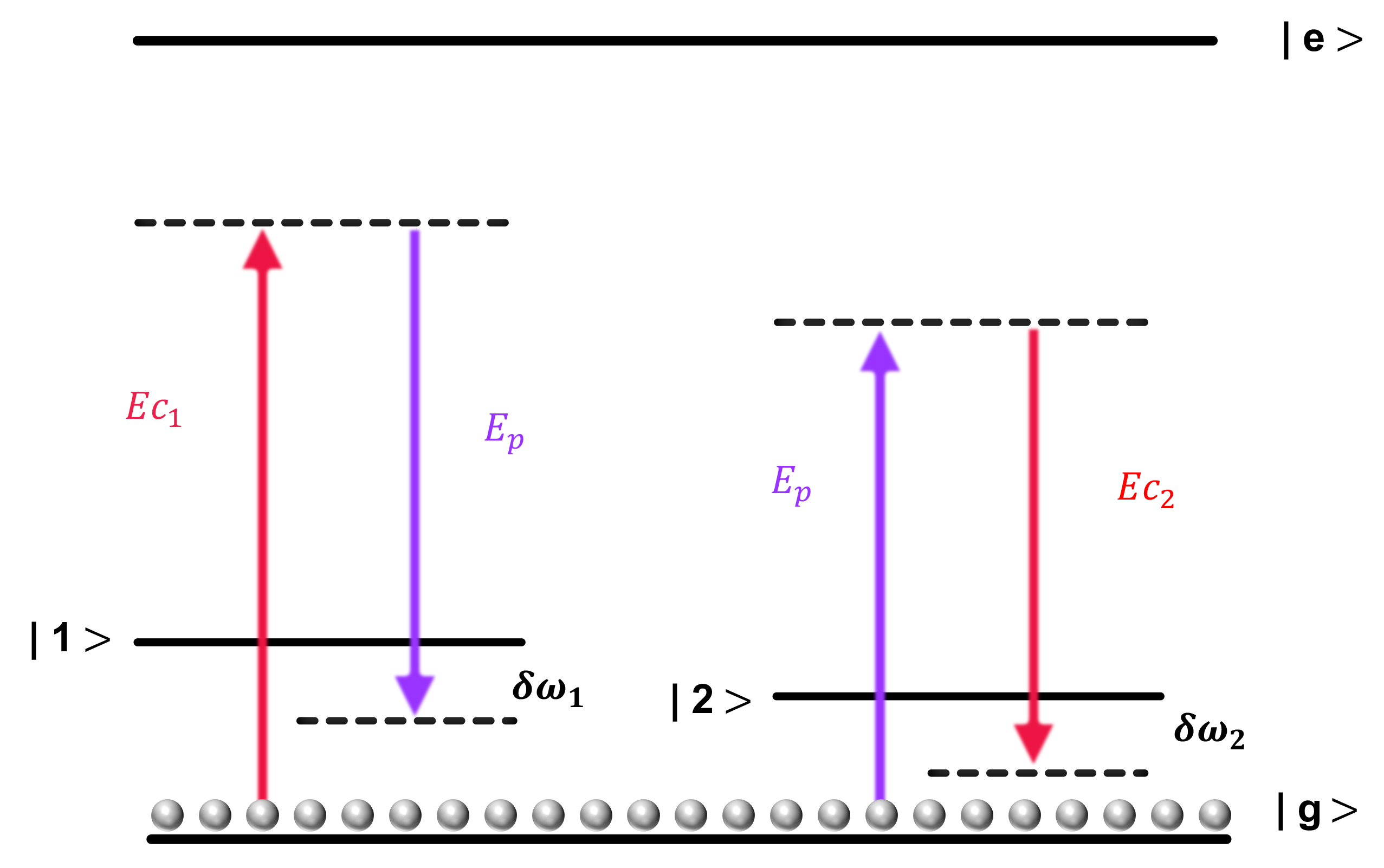}
	\caption{The energy level diagram. $E_{p}$ is the probe beam. $E_{c1}$ and $E_{c2}$ are two control beams. Two-photon couple the ground state $|g>$ to excited Raman states $|1>$ and $|2>$.  }
	\label{P2}
\end{figure} 

Non-absorption index enhancement is realized by constructing the coherent superposition of two two-level systems. According to the theoretical and experimental verification, it can be realized by far detuning a four-level system \cite{Zibrov:1996,Yavuz:2005,Proite:2008}. The energy level structure is shown in Fig.4. The controlling beams $E_{c1}$ and $E_{c2}$ construct two $\Lambda$-type frequency-level systems with the probe beam $E_{p}$. And the frequency detuning are $\delta\omega_{1}$ and $\delta\omega_{2}$ respectively. When $I_{c1}=I_{c2}$ and $\delta\omega_{1}=\delta\omega_{2}$ are satisfied, there is perfect destructive interference in the imaginary part of the susceptibility. And the real part of the susceptibility increased approximately linearly with intensities of the two controlling beams. Thus, by steering the intensity of the control beams, the refractive index of the medium can be greatly changed, and the characteristics of very low absorption can be realized at the same time. We will use this property to simulate the spacetime geometry.

Combined with the above content, we design the following experimental scheme. A Rubidium vapor cell was used as an experimental medium. During the experiment, two controlling beams are incident along a direction with an angle of $\theta$ with respect to the probe light. We use the space change attenuator to control the light intensity of the two controlling beams along the probe light direction in the atomic ensemble, so that they satisfy the following equation,
\begin{eqnarray}\label{intensity}
I_{c1}(z)=I_{c2} (z)=\frac{C}{4cos(\theta)}H^2 z^2,
\end{eqnarray}
where $C$ is a constant depending on the property of the atomic ensemble \cite{Yavuz:2005}.
In this way, the refractive index experienced by the probe light during its transmission along the atomic cell 
 satisfies $n_{2}$ in (9). This corresponds to the scheme II.  To realize scheme I, we need to prepare the controlling beams in a superposition state so that the atomic ensemble is in the cat state (\ref{superpositionworlds}). 

\begin{figure}[h]
	\includegraphics[width=1\columnwidth]{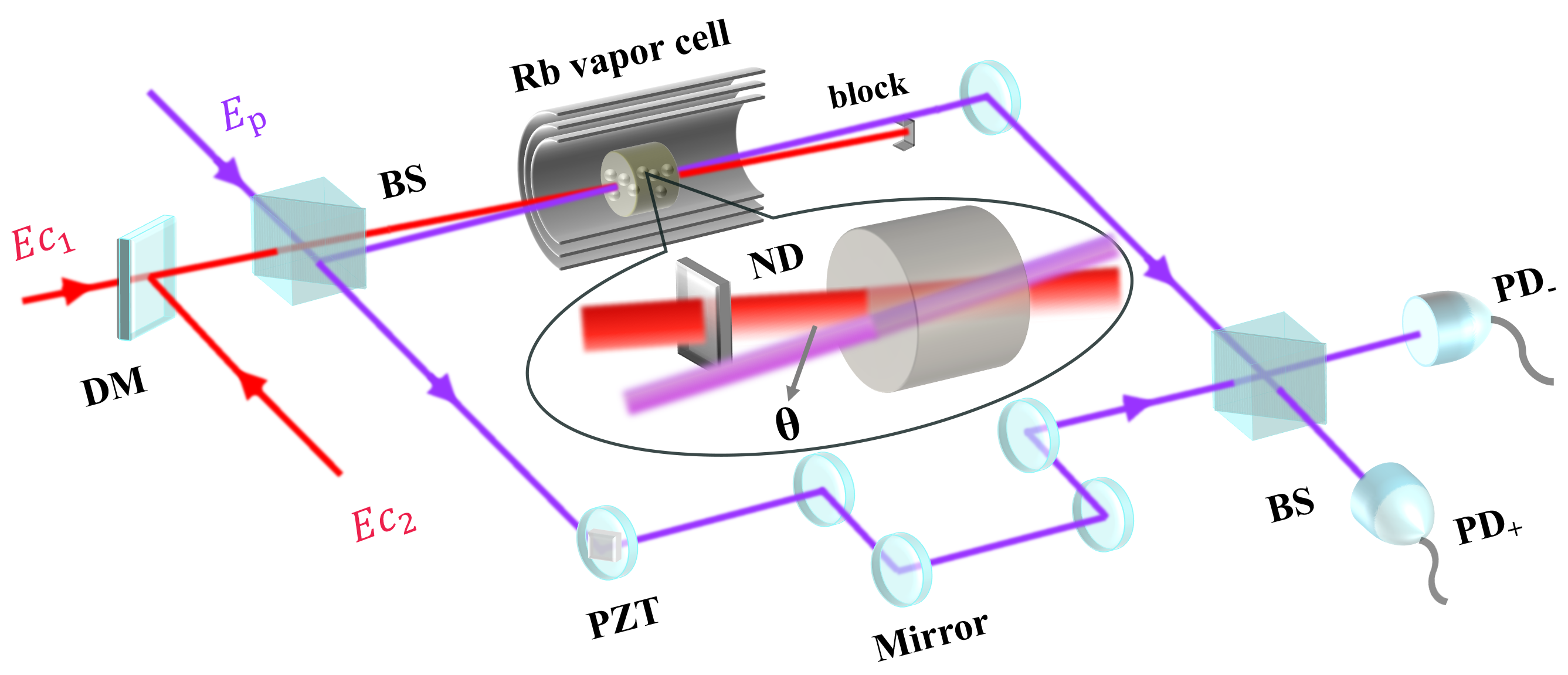}
	\caption{Schematic of experimental setup. The experimental is performed in a magnetically shielded Rb vapor cell. $E_{p}$ is the probe beam. $E_{c1}$ and $E_{c2}$ are two control beams. The inset figure shows the gradient of light intensity along the axis due to the gradient attenuator (ND) across the control beams.}
	\label{P1}
\end{figure} 

As shown in Fig.5, we put the whole device in a Mach-Zehnder interferometer (MZI). The probe light passes through the first beam splitter, one of which passes through an optical path with an atomic vapor cell, and the other passes through a path without a medium. We use the MZI to reform the information of the phase change. At the output ports $PD_{\pm}$, one  can observe  $n_0 P_{\pm}$
photons, where $n_0$ is number of probe photons and $P_{\pm}$ are the probabilities given by (\ref{probabilitiesI},\ref{probabilitiesII}). By controlling the periodic oscillation of piezoelectric ceramics, we can obtain periodic oscillation signals on the detector. 

\rrr{Conclusion} 
In this letter, we study the Schrödinger's cat state of optical parallel universes. We design two realizable experimental schemes, which enable to explore the mysterious `parallel universes'  by a MZI. The main ideas of this letter can be generalized to Hong-Ou-Mandel interference \cite{Hong:1987zz}, which can reveal the quantum entanglement of optical parallel worlds,  and to acoustic system such as Bose-Einstein condensation \cite{Anderson:1995gf,Garay:1999sk}.  Besides, our work can also help to study the quantum effect of `gravitational waves' \cite{LIGOScientific:2016aoc} in optical universes.

 \section*{Acknowledgements}
We thank Y. Wang and L. Luo for valuable comments. Zhao is supported by NSFC grant (No.11904422). Miao thank the support by NSFC grant (No.11905297).

 



\begin{thebibliography}{00}
\bibitem{note1}
Note that one can perform a further coordinate transformation $t\to t/c_1$, then the refractive indexes of table.\ref{table1} become $n\to c_1 n$, where $c_1$ is a constant. For simplicity, we set $c_1=1$ in this letter.
\bibitem{note2}
Since the atomic ensemble is in a superposition state for scheme I, there are three effective light paths in the experimental device of Fig.\ref{P1} (two effective light paths inside the vapor cell).  As a result, the probability of obtaining a photon at the output port is (\ref{probabilitiesI}) instead of $P=(1\pm\cos\Delta \phi)/2$.
\end{thebibliography}
\end{document}